\documentclass[twocolumn,showpacs,preprintnumbers,amsmath,amssymb]{revtex4}

\usepackage{graphicx}
\usepackage{dcolumn}
\usepackage{bm}

\def\R{\partial}
\def\th{\theta}
\def\TH{\Theta}
\def\ep{\varepsilon}
\def\ph{\varphi}
\def\Ps{\varPsi}

\begin{document}


\title{Dielectric anomaly in coupled rotor systems}

\author{Hiroyuki Shima and Tsuneyoshi Nakayama}
\affiliation{Department of Applied Physics, Hokkaido University, Sapporo 060-8628, Japan}


%
%

\begin{abstract}

The correlated dynamics of coupled quantum rotors carrying electric dipole moment
is theoretically investigated.
The energy spectra of coupled rotors as a function of dipolar interaction energy
is analytically solved.
The calculated dielectric susceptibilities of the system 
show the peculiar temperature dependence different from 
that of isolated rotors.

\end{abstract} 


%

%

\maketitle


%
\section{Introduction}
%

With the advent of nanotechnologies,
quantum rotors have attracted much attention
in relevance to a fundamental element of molecular scale machinery
\cite{Gimzewski,Koumura,Bermudez}.
Arrays of surface mounted quantum rotors with electric dipole moments
are of particular interest because dipole-dipole interactions
can be controlled and even designed to yield specific behavior,
such as ferroelectricity.
Ordered two-dimensional arrays of dipole rotors yield either ferroelectric
or antiferroelectric ground states, depending on the lattice type,
while disordered arrays are predicted to form a glass phase
\cite{Rozen1988,Rozen1991}.

Besides technological problems,
the microscopic dynamics of quantum rotors have extensively studied
from physical and chemical interest.
The idea of quantum rotors is applicable to interstitial oxygen impurities 
in crystalline germanium,
where oxygen atoms are quantum-mechanically delocalized 
around the bond center position \cite{Artacho97}.
The rotational of oxygen impurities around the Ge-Ge axis
has been experimentally observed by phonon spectroscopy
\cite{Gienger}.
While the rotation of oxygen impurities in Ge is weakly hindered by an azimuthal potential
caused by the host lattice,
several materials are known to show a free rotation of molecules.
An example is ammonia groups in certain Hofmann clathrates
M(NH$_3$)$_2$M'(CN)$_4$-G \cite{Wegener,Kearley,Vorderwisch},
usually abbreviated as M-M'-G,
where M and M' are divalent metal ions and G is a guest molecule.
Nearly free uniaxial quantum rotation of NH$_3$
has been observed for the first time in Ni-Ni-(C$_6$D$_6$)$_2$
by inelastic neutron scattering \cite{Wegener}.
Recently, a surprising variation of the linewidth
has been observed for Ni-Ni-(C$_{12}$H$_{10}$)$_2$ \cite{Rogalsky},
which has been interpreted by a novel line broadening mechanism
based on rotor-rotor coupling \cite{Wurger}.
It is also known that the $\beta$ phase of solid methane \cite{Press}
as well as methane hydrate \cite{Gutt}
show almost free rotation of CH$_4$ molecule.
The linewidths of methane in clathrates show inhomogeneous broadening
owing to the dipolar coupling with water molecules \cite{Gutt2}.
It is therefore expected that new interesting phenomena
will be found by investigating the influence of dipolar interaction between quantum rotors.

In the present paper, we study the correlated dynamics of coupled quantum rotors
carrying electric dipole moments.
We give the exact solution of eigenvalue problem
of interacting rotors with arbitrary configurations.
It is revealed that coupled rotors show a peculiar dielectric response
at low temperatures,
which can be interpreted by taking account of the selection rule of dipolar transition
for coupled rotors.

%
\section{The Hamiltonian}
%
 \begin{figure}[bbb]
 \vspace*{-0.5cm}
 \includegraphics[height=10cm]{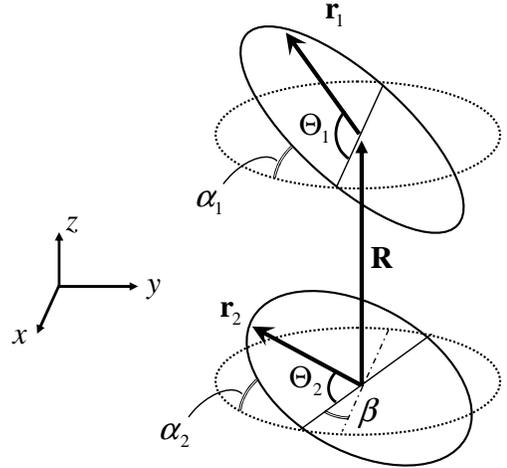}
 \vspace*{-3cm}
 \caption{Schematic configuration of coupled rotors.
 Each rotors represented by $\bm{r}_1$ and $\bm{r}_2$ rotates 
 along a ring of a radius $r$,
 and carry dipole moment $q\bm{r}_{1\,(2)}$.}
 \label{fig1}
 \end{figure}
%

Suppose two dipole rotors $q\bm{r}_1$ and $q \bm{r}_2$
separated by the vector $\bm{R}$.
The Hamiltonian for the system is given by
$H = H_K + W_D$,
where the kinetic term is 
\begin{equation}
H_K = -\frac{\hbar^2}{2I}
\left(\frac{\R^2}{\R\TH_1^2}+\frac{\R^2}{\R\TH_2^2} \right),
\label{eq_01}
\end{equation}
and the interaction term becomes
\begin{equation}
W_D = \frac{q^2}{4\pi\ep} 
     \left( \frac{1}{|\bm{R}|} + \frac{1}{|\bm{R}+\bm{r}_1-\bm{r}_2|}
           - \frac{1}{|\bm{R}+\bm{r}_1|} - \frac{1}{|\bm{R}-\bm{r}_2|} \right). 
\label{eq_02}           
\end{equation}
Here $I$ is the moment of inertia for dipole rotors
and $\ep$ the dielectric constant, respectively.
Figure 1 shows a configuration of two dipoles rotors under consideration.
We assume that rotors do not feel
any potential variation along a ring of radius $r$.
In the Jacobi coordinate, the vectors $\bm{r}_1,\bm{r}_2$ and $\bm{R}$ are given by
\begin{eqnarray}
\bm{r}_1 &=& r(\cos\TH_1,\; \sin\TH_1 \cos\alpha_1,\; \sin\TH_1 \sin\alpha_1), \nonumber \\
\bm{r}_2 &=& r(\cos\TH_2 \cos\beta - \sin\TH_2 \cos\alpha_2 \sin\beta,\; \nonumber \\
         & &\quad  \cos\TH_2 \sin\beta + \sin\TH_2 \cos\alpha_2 \cos\beta,\;
               \sin\TH_2 \sin\alpha_2), \nonumber \\
\bm{R} &=& R(0,\; 0,\; 1).
\label{eq02}
\end{eqnarray}
A spatial profile of $W_D$ as a function of $(\TH_1, \TH_2)$ is displayed in Fig. 2
by a contour plot, in which
the angles $(\alpha_1, \alpha_2, \beta)$ are set as $(\pi/4, -\pi/6, \pi/3)$.
We should remark that two minima (dark regions) and two maxima (white ones)
are located at the anti-parallel or parallel dipolar configuration,
indicating that the dipoles prefer an anti-parallel configuration.
The two minima of $W_D(\TH_1, \TH_2)$ arise from
the dipole interaction between two rotors, {\it i.e.,} 
the dipole interaction plays a key role for creating barriers and two potential minima,
which strongly affect the energy spectra and the dielectric response of the system.
\begin{figure}[ttb]
\vspace*{5.0cm}
\caption{Contour plot of the interaction term $W_D$ in $(\TH_1, \TH_2)$ plane.
Two maxima (white regions) and two minima (dark regions) are realized at positions
with differences $\Delta\TH_1\approx \pi$ and $\Delta\TH_2\approx \pi$.
Parameter values are given in the text.}
\label{fig2}
\end{figure}

Provided that the spacing $R$ is large enough 
compared with the radius $r$,
the interaction term $W_D$ can be expanded in terms of $1/R$.
The lowest-order term has the form of a dipolar interaction given by 
\begin{equation}
W_D^{(3)} = \frac{q^2}{4\pi\ep R^3}\left\{
 \bm{r}_1\cdot\bm{r}_2
-\frac{3(\bm{r}_1\cdot\bm{R})(\bm{r}_2\cdot\bm{R})}{R^2}
\right\}.
\label{eq_04}
\end{equation}
The higher-order term $W_D^{(4)} \equiv W_D-W_D^{(3)} $ 
is of the order of $O(r^3/R^4)$,
which can be negligible for the case $R\gg r$.
Actually we have confirmed that the calculated results presented in this paper
change very little by taking into account the term $W_D^{(4)}$.

%
\section{Eigenvalues and eigenfunctions}
%

The Schr\"odinger equation for the Hamiltonian
$H_0 = H_K + W_D^{(3)}$ has analytic solutions as shown below.
Transforming variables to
$ \th_1=(\TH_1+\TH_2)/2$ and $\th_2=(\TH_1-\TH_2)/2$,
%
%
Eqs. (\ref{eq_01}) and (\ref{eq_04}) yields
\begin{eqnarray}
H_K &=& -\frac{\hbar^2}{4I}\left(\frac{\R^2}{\R\th_1^2}+\frac{\R^2}{\R\th_2^2} \right), 
\label{eq_05} \\
W_D^{(3)} &=& \frac{q^2 r^2}{4\pi \ep R^3} \sum_{i=1}^2 c_i \cos2(\th_i+\gamma_i).
\label{eq_06}
\end{eqnarray}
The parameters $c_i$ and $\gamma_i$ $(i=1,2)$ are functions of 
angles $\alpha_1, \alpha_2$ and $\beta$ defined in Fig. 1,
whose explicit forms are given by
\begin{equation}
c_i = \frac12 \sqrt{x_i^2+y_i^2}, \quad 
\gamma_i = \frac 12 \tan^{-1}\left( \frac{-x_i}{y_i} \right),
\label{eq_09}
\end{equation}
%
%
%
with the definitions
\begin{eqnarray}
x_1 &=& \sin\beta ( \cos\alpha_1 - \cos\alpha_2 ), \nonumber \\
x_2 &=& \sin\beta ( \cos\alpha_1 + \cos\alpha_2 ), \nonumber \\
y_1 &=& \cos\beta( 1 - \cos\alpha_1 \cos\alpha_2 )
      +2 \sin\alpha_1 \sin\alpha_2, \nonumber \\
y_2 &=& \cos\beta (1 + \cos\alpha_1 \cos\alpha_2)
      -2 \sin\alpha_1 \sin\alpha_2.
\label{eq_10}
\end{eqnarray}
Consequently, we can decompose the Schr\"odinger equation 
$H_0\Ps_0(\th_1, \th_2)=E_0\Ps_0(\th_1, \th_2)$ 
into two independent Mathieu equations.
Setting 
$\Ps_0(\th_1,\th_2) = \ph_1(\th_1)\ph_2(\th_2)$, we obtain
\begin{equation}
-\frac{\R^2\ph_i}{\R\th_i^2} +\frac{2}{E_K}
\left[c_i E_D \cos 2(\th_i+\gamma_i)-E_i \right]\ph_i = 0, \quad [i=1,2]
\label{eq_09}
\end{equation}
where the quantities $E_K=\hbar^2/(2I)$ and $E_D=q^2 r^2/(4\pi\ep R^3)$ 
represent the kinetic and interaction energy, respectively.
The eigenvalue $E$ of the initial Schr\"odinger equation is expressed as the sum of
$E=E_1+E_2$.
Note that the periodic terms $\propto \cos2(\theta_i+\gamma_i)$
originate from two minima (or maxima) of the interaction term
$W_D(\TH_1,\TH_2)$ shown in Fig.~2 \cite{foot1}.

Eigenfunctions of Eq.(\ref{eq_09}) are described by 
four types of the Mathieu functions, given by ce$_{2n}(v_i, \th_i)$,
se$_{2n+1}(v_i, \th_i)$, ce$_{2n+1}(v_i, \th_i)$ and 
se$_{2n+2}(v_i, \th_i)$ with the definitions
$v_i \equiv c_i E_D/E_K$ and $n=0,1,2\cdots$. 
Each of them belongs to a different eigenvalue
and can be expressed in terms of the Fourier-cosine expansion;
for instance,
\begin{equation}
{\rm ce}_{2n}(v_i,\th_i) = \sum_{m=0}^{\infty} A_{2m}^{(2n)}(v_i)\cos 2m(\th_i+\gamma_i).
\label{eq05}
\end{equation}
The coefficients $\{A_{2m}^{(2n)}\}$ are determined by a successive relation
obtained by substituting Eq.(\ref{eq05}) into Eq.(\ref{eq_09}).
The amplitudes of $\{A_{2m}^{(2n)}\}$ rapidly decrease with increasing $m$,
so that we can truncate the summation in Eq.~(\ref{eq05}) at $m=20$ in actual calculations.
 \begin{figure}[ttb]
 \includegraphics[height=7.5cm]{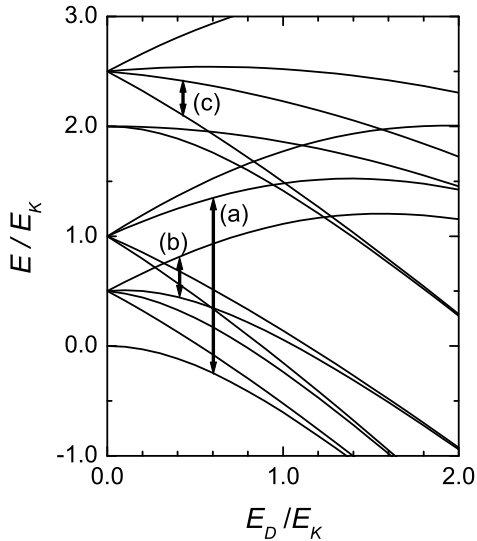}
 \caption{The energy spectra of the paired-rotor as a function of $E_D$.
 Solid arrows indicate a part of allowed dipole transitions
 for the component $p_x$ in the case of 
 $(\alpha_1, \alpha_2, \beta)=(\pi/4,0,0)$.
 The explanation on three labels (a)-(c) is given in text.}
 \label{fig2}
 \end{figure}

Figure 3 plots the calculated spectra of eigenenergies $E=E_1+E_2$ as a function of $E_D$,
where $E_K$ is taken as an energy unit.
The angles $(\alpha_1,\alpha_2,\beta)$ are set to be $(\pi/4,0,0)$
as an example.
We find, though some levels are degenerate when $E_D=0$,
they split off for finite $E_D$ with a monotonous variation with increasing $E_D$.
For high-$E_D$ limit, some levels become degenerate again.
It  indicates that the relative motion of paired-rotors is frozen out for 
$E_D\gg E_K$ due to the strong Coulomb interaction. 
This behavior can be understood
from the spatial profile of the interaction term $W_D(\TH_1, \TH_2)$
shown in Fig. 2.
With increasing $E_D$, the depths of two minima of $W_D(\TH_1,\TH_2)$
grow, and larger barrier-heights hinder the quantum transition of a particle
through the barrier. 
This gives rise to localized wavefunctions around these minima.
Consequently, in the limit of $E_D \gg E_K$,
the amplitude of the eigenfunctions are strongly localized around two minima,
and these two localized eigenstates are nearly degenerate.
Even if the higher-order term $W_D^{(4)}$ is taken into account,
the energy spectra does not change much,
since it only slightly disturbs the symmetry of the depths of two minima
shown in Fig. 2.
When varying the angles $(\alpha_1,\alpha_2,\beta)$,
the curves in Fig. 3 slightly shift to upwards and/or downwards
except for the unchanged values of $E$ at $E_D=0$.

%
\section{Dielectric susceptibilities}
%

%
%
%
%
%
%
%

Let us consider the dielectric response of dipole rotors
coupled via dipolar interaction.
The real part of 
the frequency-dependent dielectric susceptibility is expressed as
\begin{eqnarray}
\chi_{\mu\mu}(\omega,T) &=&
-\frac{2}{\ep Z}\sum_{j,l\ne j}
\left| \langle E_j|p_{\mu}|E_l \rangle \right|^2 \nonumber \\
&\times&
 \frac{E_j-E_l}{(E_j-E_l)^2-(\hbar\omega)^2} \exp\left(-\frac{E_j}{k_B T}\right),
 \label{eq08}
\end{eqnarray}
where $Z=\sum_j \exp(-E_j/k_B T)$ is the partition function,
and $|E_j \rangle$ is the eigenvector belonging to the eigenvalue $E_j$.
The quantity $p_{\mu}$ is the $\mu$-component 
of the total dipole moment $\bm{p}=q(\bm{r}_1 + \bm{r}_2)$,
which depend on the relative orientation with respect to the external field.
We should note that the susceptibility depends on 
the selection rules for dipole transitions
between different eigenstates.
In Fig.~3, allowed dipole transitions for $p_x$
are indicated in part by solid arrows.
Note that only a part of allowed transitions are shown in the figure,
which are dominant for the dielectric susceptibility 
$\chi(\omega,\,T)$ at temperatures $T \approx E_K/k_B$.
The rest of allowed dipolar transitions do not contribute 
to the susceptibility given by Eq.(\ref{eq08}),
because the energy difference $|E_j-E_l|$ is so large 
and/or the Boltzmann factor $\exp(-E_j/k_B T)$ become much smaller than unity.
The interpretation on three labels (a)-(c) shown in Fig.~3 will be given later.
%
 \begin{figure}[ttb]
 \includegraphics[height=8.5cm]{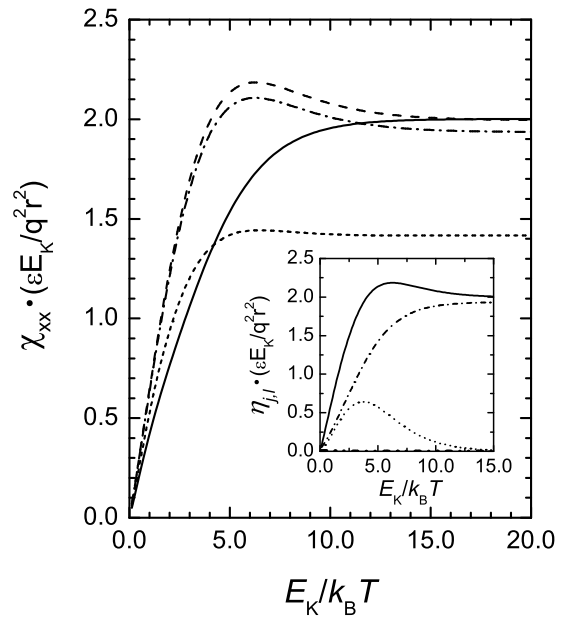}
 \caption{The dielectric susceptibility $\chi_{xx}(T)$ for the zero-frequency limit $\omega\to 0$
 as a function of the inverse temperature $1/T$.
 The strength of dipolar interaction is increased 
 from top to bottom; i) $E_D=0$ (solid), ii) $E_D = 0.01$ (dashed), 
 iii) $E_D=0.1$ (dash-dotted), and iv) $E_D=1.0$ (dotted) in units of $E_K$.
 Bumps at around $E_K/(k_B T) \approx 5.0$ appear in the cases of ii) and iii).
 Inset shows three components of $\eta_{j,l}(T)$ for the case of iii), 
 whose definitions are given in text.}
 \label{fig3}
 \end{figure}

We have calculated the temperature dependence of the dielectric susceptibility
$\chi_{\mu\mu}(\omega, T)$ for various $E_D$.
Figure 4 shows the calculated results of {\it dc} susceptibility $\chi_{xx}(T)$
normalized by a factor $q^2 r^2/(\varepsilon E_K)$.
We have taken four values of $E_D/E_K$;
the solid line ($E_D=0$), the dashed one ($E_D/E_K=0.01$),
the dash-dotted one ($E_D/E_K=0.1$), and the dotted one ($E_D/E_K=1.0$).
The angles are set to be $(\alpha_1,\alpha_2,\beta)=(\pi/4,0,0)$ for all $E_D$.
For the case of $E_D=0$, the susceptibility monotonically increases
with decreasing temperature, and becomes constant at lower temperatures.
The crossover temperature between the steady increase and the almost constant value in Fig. 4
is determined by the minimum-energy difference of eigenstates at $E_D=0$
{\it that are allowed for dipole transition}, namely, indicated as (a) in Fig.~3.
For the case of $E_D/E_K > 1$,
the strong Coulomb interaction prevents from the relative motion of rotors
so that the magnitude of the susceptibility $\chi(T)$ decreases with increasing $E_D$.

It is noteworthy that, for relatively weak interaction $E_D/E_K<0.1$,
a bump is appeared in the susceptibility at about $E_K/(k_B T) \approx 5.0$.
The kinetic energy $E_K=\hbar^2/(2I)$ for actual rotating molecules
is the order of 1 meV \cite{Gutt},
indicating that the characteristic temperature 
$T^* =E_K/k_B \times 0.2$ corresponding to the bump
is estimated as about 1 K.
We made sure that the bump can be observed 
for any angles $(\alpha_1,\;\alpha_2,\; \beta)$ when $E_D/E_K$ is less than $0.1$.
This anomaly stems from the correlated rotation of paired-rotors 
via the dipolar interaction,
and can be interpreted by the argument on the selection rule for dipolar transition.

To understand the origin of the bump,
we decompose the total susceptibility $\chi_{xx}(T)$ 
give in Eq.(\ref{eq08}) as 
\begin{eqnarray}
\chi_{xx}(T) &=& \sum_{(j,l\ne j)} \eta_{j,l}(T), \label{eq_star} \\
\eta_{j,l}(T) &=&
-\frac{2}{\ep Z}
\frac{\left| \langle E_j|p_x|E_l \rangle \right|^2}{E_j-E_l} \nonumber \\
&\times& \left[ \exp\left(-\frac{E_j}{k_B T}\right)-\exp\left(-\frac{E_l}{k_B T}\right) \right],
\label{eq09}
\end{eqnarray}
where $\sum_{(j,l\ne j)}$ is the summation over all possible combinations of $(j,l)$
under the condition $l \ne j$.
Note the fact that only {\it three} components of $\eta_{j,l}(T)$ are responsible 
for the total susceptibility (\ref{eq_star})
around the characteristic temperature $T^*$.
We denote those components by $\eta^a$, $\eta^b$, and $\eta^{c}$,
which are characterized by the eigenfunction 
$\Ps_j=\langle \th_1, \th_2 | E_j \rangle$ and $\Ps_l=\langle \th_1, \th_2 | E_l \rangle$ 
as follows;
\begin{eqnarray}
\eta^a ;&\;& \Ps_j={\rm ce}_0(\th_1) {\rm ce}_0(\th_2),\;\; \Ps_l
={\rm ce}_1(\th_1) {\rm ce}_1(\th_2), \label{eq10} \\
\eta^b ;&\;& \Ps_j={\rm ce}_0(\th_1) {\rm ce}_1(\th_2),\;\; \Ps_l
={\rm ce}_1(\th_1) {\rm ce}_0(\th_2), \label{eq11} \\
\eta^c ;&\;& \Ps_j={\rm se}_1(\th_1) {\rm se}_2(\th_2),\;\; \Ps_l
={\rm se}_2(\th_1) {\rm se}_1(\th_2). \label{eq12} 
\end{eqnarray}
The alphabets subscribed on $\eta$ correspond to three dipolar transitions
labeled by (a)-(c) shown in Fig.~3.
For example, the solid arrow of (a) in Fig.~3 
connects the eigenstates $\Ps_j$ and $\Ps_l$
defined in Eq.(\ref{eq10}).

For weak coupling $E_D\ll E_K$, the solution of the Mathieu equation (9) is easily solved.
In the lowest order of the perturbation theory,
the eigenvalues $E_i (i=1,2)$ read in
\begin{equation}
E_i = \frac{E_K}{2}n^2 + a E_D \quad (n=0,\pm 1, \pm 2,\cdots)
\label{eq_add1}
\end{equation}
with a constant $a$.
The solution (\ref{eq_add1}) gives the eigenenergies of the states 
$|E_j \rangle$ and $| E_l \rangle$ relevant to the three components as follows;
\begin{eqnarray}
\eta^a; \; \quad (E_j,E_l) &=& (0, E_K+\delta E_a), \label{eq_damie} \\
\eta^b; \; \quad (E_j,E_l) &=& (\frac{E_K}{2} \pm \delta E_b, \frac{E_K}{2} \mp \delta E_b), \label{eq_niju} \\
\eta^c; \; \quad (E_j,E_l) &=& (\frac{5E_K}{2} \pm \delta E_c, \frac{5E_K}{2} \mp \delta E_c).
\end{eqnarray}
The small corrections $\delta E$ stem from the small interaction energy $E_D\ll E_K$.
Substituting these eigenenergies into Eq. (13),
we find that the three components are approximated by
\begin{eqnarray}
\eta^a (u) &=& \frac{2 p_a^2}{\ep E_K} \cdot \frac{1-e^{-u}}{Z(u)},  \\
\eta^b (u) &=& \frac{2 p_b^2}{\ep E_K} \cdot \frac{u e^{-u/2}}{Z(u)}, \\
\eta^c (u) &=& \frac{2 p_c^2}{\ep E_K} \cdot \frac{u e^{-5u/2}}{Z(u)},
\end{eqnarray}
where we defined $u=E_K/(k_B T)$.
The quantities $p_a$, $p_b$ and $p_c$ equal to the value of 
$\left| \langle E_j|p_x|E_l \rangle \right|$
for the case of $\eta^a$, $\eta^b$ and $\eta^c$, respectively.
The explicit form of the partition function $Z(u)$ is
\begin{eqnarray}
Z(u)=1 &+& 4e^{-u/2}+4e^{-u}+4e^{-2u}+8e^{-5u/2} \nonumber \\
 &+& 4e^{-4u}+4e^{-9u/2}+8e^{-5u}+ \cdots,
\end{eqnarray}
which monotonically decreases with rasing $u$ and reaches unit for the limit $u \to \infty$.
This means that the component $\eta^a(u)$ is a monotonic increase function of $u$.
On the other hand, the components $\eta^b(u)$ and $\eta^c(u)$ is convex functions
giving a maximum at finite $u$.
The conditions of $u$ for the maximum of $\eta^b$ and $\eta^c$
are expressed by
\begin{eqnarray}
1-\frac{u}{2}-u\frac{Z'(u)}{Z(u)} &=& 0, \quad \mbox{for } \eta^b, \label{eq_ad2} \\
1-\frac{5}{2}u-u\frac{Z'(u)}{Z(u)} &=& 0, \quad \mbox{for } \eta^c \label{eq_ad3}.
\end{eqnarray}
The solutions of the Eqs. (\ref{eq_ad2}-\ref{eq_ad3}) is estimated as
$u\approx 4$ for $\eta^b$ and $u\approx 0.5$ for $\eta^c$.
Since the total susceptibility $\chi(T)$ is given by the summation $\eta^a+\eta^b+\eta^c$,
it is expected that the convex features of $\eta^b(u)$ and $\eta^c(u)$ cause in the bump of 
the total susceptibility at $u\approx 5$ shown in Fig. 4.

The argument is clarified by the numerical results shown in the inset of Fig. 4,
where the $u$-dependence of the components for $E_D/E_K=0.1$
are displayed; $\eta^a$ (dashed-dotted), $\eta^b$ (dotted), $\eta^c$ (dashed-dotted-dotted),
together with that of the total susceptibility $\chi=\eta^a+\eta^b+\eta^c$ (solid).
The component $\eta^b$ clearly shows a maximum at $u\approx 4$,
whereas the contribution of $\eta^c$ is negligible due to the factor $e^{-5u/2}$
in Eq. (23).
As a result, the summation $\eta^a(u)+\eta^b(u)$ shows a bump at $u=5.0$,
which is the origin of the anomalous bump of the total susceptibility $\chi(T)$
at the characteristic temperature $T^*=E_K/k_B \times 0.2$.
We should note here that, if quantum rotors are not interacting at all,
the component $\eta^b$ exactly vanished due to the degeneracy $E_j=E_l=E_K/2$
 (See Eq. (\ref{eq_niju}))
and only the component $\eta^a$ is dominant for the total susceptibility $\chi(T)$.
This means that the total susceptibility is a monotonic function as the same as $\eta^a$
so that the bump does not emerge.
The anomalous bump of the susceptibility,
therefore, manifests the relevance of the dipolar interaction
to the dielectric response of quantum rotors.

%
\section{Conclusions}
%

%
%

It is important to recall experiments reported in Ref. \cite{Enss2},
for the dielectric susceptibility of KCl crystals with Li defects.
It has been found that the susceptibility does not scale linearing with the Li concentration,
and even becomes smaller with increasing the concentration ($\approx 1000$ ppm), 
where the interaction between defects becomes relevant.
In addition, a bump of the susceptibility is observed at about 200 mK for concentrations
of 200-1000 ppm. 
These temperature dependences of the susceptibility together with the bumps
are recovered well by our results shown in Fig.~4.
Noting that defects in both systems move along closed loops and correlated each other,
it is natural to assume that the similar picture holds.
For a quantitative discussion, of course,
one should take into account the effect of potential variation
hindering the free rotation of Li$^+$, which is caused by the Coulomb interaction
between a mobile Li$^+$ ion and the host atoms K$^+$ and Cl$^-$.
The problem has been theoretically investigated in Ref. \cite{Wurger2, Wurger3}
based on the two-level tunneling model.

In conclusion, we have investigated the quantum dynamics
of two dipole rotors coupled via dipolar interaction.
By solving analytically the eigenvalue problem of coupled rotors,
we have demonstrated the energy spectra of coupled rotors 
as a function of dipolar interaction.
The anomalous temperature dependence of dielectric susceptibility
is also shown.
Our model is so general that it should be applicable
in a variety of physical context relevant to quantum rotors.

%
%

\begin{acknowledgments}

One of the authors (H.S) was financially supported in part 
by the NOASTEC Foundation for young scientists.
This work was supported in part by a Grant--in--Aid for Scientific
Research from the Japan Ministry of Education, Science, Sports and Culture.

\end{acknowledgments}

%
%


\begin{thebibliography}{99}
\bibitem{foot1} The Mathieu equation was employed in Ref. \cite{Kettemann}
for calculating eigenenergies of single quantum rotor.
The main difference between the Mathieu equation used in
Ref. \cite{Kettemann} and that in this
paper is the physical origin of two potential minima expressed by the terms 
$\propto \cos(2\theta)$ in Eq. (\ref{eq_09}) .
The model of Ref. \cite{Kettemann} does not take into account the interactions
between rotors, namely, two potential minima  for single rotors are presumed.
Equation (\ref{eq_09}) was derived as a result
of the dipole interaction, 
so the similarity is fortuitous, having no underlying physical origin.
\end{thebibliography}
\end{document}